\documentclass[showpacs,preprintnumbers,prl,a4paper,10pt,twocolumn]{revtex4}%
\usepackage{amssymb}
\usepackage{amsmath}
\usepackage{graphicx}
\usepackage{dcolumn}
\usepackage{bm}
\usepackage{amsfonts}%
\providecommand{\U}[1]{\protect\rule{.1in}{.1in}}
\begin{document}
\title{Prospects of employing superconducting stripline resonators for studying the
dynamical Casimir effect experimentally}
\author{Eran \surname{Arbel-Segev}}
\email{segeve@tx.technion.ac.il}
\author{Baleegh Abdo}
\author{Oleg Shtempluck}
\author{Eyal Buks}
\affiliation{Department of Electrical Engineering, Technion, Haifa 32000 Israel.}
\author{Bernard Yurke}
\affiliation{Bell Laboratories, Lucent Technologies, 600 Mountain Avenue, Murray Hill, NJ 07974}
\date{\today }

\begin{abstract}
We discuss the prospects of employing an NbN superconducting microwave
stripline resonator for studying the dynamical Casimir effect experimentally.
Preliminary experimental results, in which optical illumination is employed
for modulating the resonance frequencies of the resonator, show that such a
system is highly promising for this purpose. Moreover, we discuss the
undesirable effect of heating which results from the optical illumination, and
show that degradation in noise properties can be minimized by employing an
appropriate design.

\end{abstract}

\maketitle

The term dynamical Casimir effect (DCE) refers usually to the problem of an
electromagnetic (EM) cavity with periodically moving walls. The quantum theory
of electrodynamics predicts that under appropriate conditions, photons should
be created in such a cavity out of the vacuum fluctuations \cite{DCE_Moore70}.
Such motion-induced radiation is closely related to the Unruh - Davies effect,
which predicts that an observer of the EM field in a uniformly accelerating
frame would measure thermal radiation with an effective temperature given by
$\hbar a/2\pi k_{B}c$, where $a$ is the acceleration, $k_{B}$ is the
Boltzmann's constant, and $c$ is the light velocity in vacuum. \ Moreover, the
equivalence principle of general relativity relates the later effect with the
so-called Hawking radiation of black holes
\cite{DCE_Fulling76,DCE_Dewitt75,DCE_Hawking74,DCE_Rosu01,DCE_Yurke87}.

Efficient production of photons can be achieved by employing parametric
resonance conditions \cite{DCE_Dodonov96}. Consider the case where the cavity
walls oscillate at twice the resonance frequency of one of the cavity modes
(primary parametric resonance). In this case the angular resonance frequency
$\omega_{r}$ varies in time according to
\begin{equation}
\omega_{r}(t)=\omega_{0}\left[  1+\xi\cos\left(  2\omega_{0}t\right)  \right]
. \label{omega r}%
\end{equation}
The system's response to such an excitation depends on the dimensionless
parameter $\xi Q$, where $Q$ is the quality factor of the resonator
\cite{ParAmp_Landau60}. When $\xi Q<1$, the system is said to be in the
sub-threshold region, while above threshold, when $\xi Q>1$, the system breaks
into oscillations. Achieving the condition $\xi Q>1$ requires that the shift
in the resonance frequency exceeds the width of its peak.

So far the DCE has not been verified experimentally
\cite{DCE_Braggio05,Casimir_Woo-Joong06}. It turns out that creation of
photons in the case of a cavity with moving walls requires that the peak
velocity of the moving walls must be made comparable to light velocity, a task
which is extremely difficult experimentally \cite{DCE_Dodonov01}. When this is
not the case the system is said to be in the adiabatic regime, where the
thermal average number of photons is time independent.

An alternative method for realizing the DCE was pointed out by Yablonovitch
\cite{DCE_Yablonovitch89}, who proposed that modulating the dielectric
properties of a material in an EM cavity might be equivalent to moving its
walls. As a particular example, he considered the case of modulating the
dielectric constant $\epsilon$ of a semiconductor by optical pulses that
create electron-hole pairs. The modulation frequency achieved by this method
is limited by the recombination time of electron-hole pairs, which can be
relatively fast in some semiconductors \cite{DCE_Lozovik95}. Based on these
ideas, a novel experimental approach for the detection of the DCE was recently
proposed \cite{DCE_Braggio05}. However, implementing this approach might be
very difficult \cite{DCE_Dodonov05}. The change in the dielectric constant
$\Delta\epsilon=\Delta\epsilon^{\prime}+i\Delta\epsilon^{\prime\prime}$ that
occurs due to creation of electron-hole pairs in a semiconductor can be found
by employing the Drude model \cite{Ashcroft76}. In the microwave region one
finds $\Delta\epsilon^{\prime}/\Delta\epsilon^{\prime\prime}\simeq\omega\tau$
(see Eq. (1.35) of Ref. \cite{Ashcroft76}), where $\tau$ is the momentum
relaxation time, and $\omega$ is the angular frequency. However, for all known
semiconductors in the microwave region $\omega\tau\ll1$, and consequently,
unless the resonator is carefully designed, exciting charge carriers will
mainly lead to undesirable broadening of the resonance peaks while the
frequency shift is expected to be relatively small.

On the other hand, parametric excitation by modulating $\epsilon$ can be
implemented with a superconductor instead of a semiconductor. Optical
radiation in the latter case allows modulation of the relative density of
superconducting electrons and that of normal electrons.\ The resultant change
in the dielectric constant $\Delta\epsilon$, which can be found from the
two-fluid model \cite{SupCond_Orlando91}, depends on the ratio between the
London length $\lambda$ and the skin depth of normal electrons $\delta$.
According to London's theory the ratio between these length scales is given by
$\lambda_{0}/\delta=\left(  \omega\tau/2\right)  ^{1/2}$, where $\lambda_{0}$
is the London length at zero temperature (see Eqs. (14.21) and (34.9) of Ref.
\cite{Ashcroft76}). Consequently, in the microwave region one finds
$\Delta\epsilon^{\prime}/\Delta\epsilon^{\prime\prime}\simeq1/\omega\tau\gg1$
\cite{supRes_Xu96}. This property of superconductors significantly facilitates
achieving parametric gain by optically modulating $\epsilon$ since frequency
shift can be made much larger in comparison with an undesirable peak
broadening (see Eq. (A4) in Ref. \cite{supRes_Golosovsky95}). Indeed,
quasi-static resonance frequency shift by optical radiation
\cite{supRes_Cho03,supRes_Tsindlekht94}, or high-energy particles
\cite{supRes_Day03} (for which the required condition $\xi Q\cong1$ has been
achieved) has been demonstrated.

In the present paper we discuss the prospects of employing a superconducting
stripline resonator for experimentally studying the DCE. Our preliminary
results show that such a system is highly promising for this purpose. In
particular we show that achieving parametric gain by optically modulating the
resonance frequency is feasible. We employ a novel configuration in which a
hot electron detector (HED) is implemented as an integrated part of a
microwave superconducting NbN stripline resonator. An optical pulse impinging
on the HED results in a generation of hotspots, namely, small sections of the
HED are heated above the superconducting critical temperature and thus undergo
a transition from the superconducting phase to the normal one
\cite{HED_Kadin96}. As a result the impedance of the HED is substantially
modified \cite{supRes_Saeedkia05}. Both the change in the resistive and
inductive parts of the HED impedance contributes to the resonance shift
\cite{supRes_Day03,Segev06a}, whereas the undesirable change in the damping
rate is relatively small. In general however, employing such a modulation
scheme results in some undesirable heating of the illuminated superconductor;
whereas the quantum nature of the DCE requires operating at vary low
temperatures. In the last part of this paper we discuss theoretically the
expected effect of such heating on the noise properties of the system. In
particular, we study the conditions for achieving noise squeezing when
homodyne detection scheme is employed for readout. Our results show that the
undesirable effect of heating can be minimized by employing an appropriate
design. Note that, noise squeezing, which is expected to occur in the
sub-threshold region, bares the same underlying physics as the DCE in the
overcritical region
\cite{DCE_Dodonov90,DCE_Dodonov99,DCE_Lozovik95,DCE_Plunien00}.

\begin{figure}
[ptb]
\begin{center}
\includegraphics[
height=6.7037cm,
width=8.2874cm
]%
{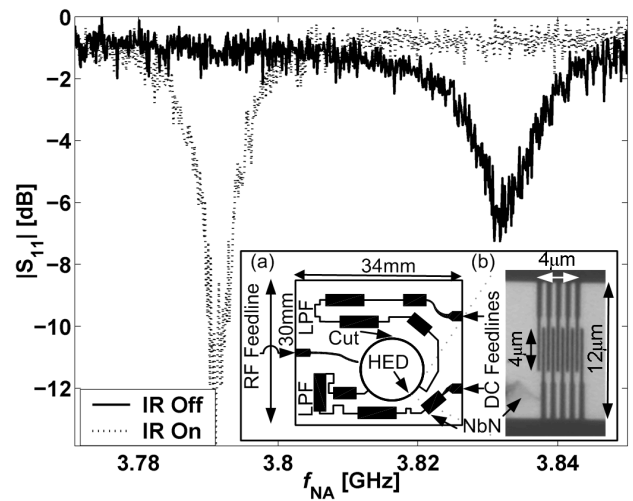}%
\caption{Reflection coefficient ($|S_{11}|$) measurement in the vicinity of
the second resonance mode with (dotted) and without (solid) IR laser
illuminating the HED, while a sub-critical dc current is applied through it.
The inset shows the layout of the device and an optical image of the HED.}%
\label{reflection}%
\end{center}
\end{figure}

The design of the resonator takes advantage of recent progress in the field of
superconducting single photon detectors. Switching time in superconductors is
usually limited by the relaxation process of high-energy quasi-particles, also
called 'hot-electrons', giving their energy to the lattice, and recombining to
form Cooper pairs. Recent experiments with such photodetectors have
demonstrated an intrinsic switching time on the order of $30%
%TCIMACRO{\unit{ps}}%
%BeginExpansion
\operatorname{ps}%
%EndExpansion
$ and a counting rate exceeding $2%
%TCIMACRO{\unit{GHz}}%
%BeginExpansion
\operatorname{GHz}%
%EndExpansion
$ (see \cite{HED_Goltsman05} and references therein). The circuit layout is
illustrated in inset $(\mathrm{a})$ of Fig. \ref{reflection}. The resonator is
designed as a stripline ring, having a characteristic impedance of $50%
%TCIMACRO{\unit{\U{3a9}}}%
%BeginExpansion
\operatorname{\Omega }%
%EndExpansion
$. It is composed of $8$-nm-think NbN film deposited on a Sapphire wafer. The
first few resonance frequencies fall within the range of $2-8%
%TCIMACRO{\unit{GHz}}%
%BeginExpansion
\operatorname{GHz}%
%EndExpansion
$. A feedline, weakly coupled to the resonator, is employed for delivering the
input and output signals. A HED is integrated into the structure of the ring.
Its angular location, relative to the feedline coupling location, maximizes
the RF current amplitude flowing through it in one of the resonance modes, and
thus maximizes its coupling to that mode. The HED has a meander shape (inset
$(\mathrm{b})$ of Fig. \ref{reflection}) that consists of nine strips. Each
strip has a characteristic area of $0.15\times4%
%TCIMACRO{\unit{\U{3bc}m}}%
%BeginExpansion
\operatorname{\mu m}%
%EndExpansion
^{2}$ and the strips are separated one from another by approximately $0.25%
%TCIMACRO{\unit{\U{3bc}m}}%
%BeginExpansion
\operatorname{\mu m}%
%EndExpansion
$ \cite{HED_Zhang03}. The HED operating point can be maintained by applying dc
bias. The dc\ bias lines are designed as two superconducting on-chip low-pass
filters (LPF). A cut of $20%
%TCIMACRO{\unit{\U{3bc}m}}%
%BeginExpansion
\operatorname{\mu m}%
%EndExpansion
$ is made in the perimeter of the resonator, to force the dc bias current flow
through the HED. Further design considerations, fabrication details as well as
calculation of normal modes can be found elsewhere \cite{Segev06a}.
Measurements are carried out in a fully immersed sample in liquid Helium.%
%TCIMACRO{\FRAME{ftbpFU}{8.2874cm}{6.7037cm}{0pt}{\Qcb{Reflection coefficient
%($|S_{11}|$) measurement in the vicinity of the second resonance mode with
%(dotted) and without (solid) IR laser illuminating the HED, while a
%sub-critical dc current is applied through it. The inset shows the layout of
%the device and an optical image of the HED.}}{\Qlb{reflection}}{fig1.eps}%
%maintain-aspect-ratio TRUE;  display "ICON";  valid_file "F";
%width 8.2874cm;  height 6.7037cm;  depth 0pt;  original-width 6.7463in;
%original-height 5.4346in;  cropleft "0";  croptop "1";  cropright "1";
%cropbottom "0";  filename 'fig1.eps';file-properties "XNPEU";}}}%
%BeginExpansion
%EndExpansion

The effect of infrared (IR) laser illumination on one of the resonance
frequencies of the resonator is shown in Fig. \ref{reflection}, which exhibits
two reflection coefficient ($|S_{11}|$) measurements which were performed near
the second resonance frequency $f_{0}=3.71%
%TCIMACRO{\unit{GHz}}%
%BeginExpansion
\operatorname{GHz}%
%EndExpansion
$ and obtained using a network analyzer. The HED is biased with a sub-critical
dc current of $4.14%
%TCIMACRO{\unit{\U{3bc}A}}%
%BeginExpansion
\operatorname{\mu A}%
%EndExpansion
$, which only slightly increases the damping rate relative to the measured
damping rate in the case where the HED is unbiased (the critical dc current
while injecting the RF signal is $I_{C}=4.35%
%TCIMACRO{\unit{\U{3bc}A}}%
%BeginExpansion
\operatorname{\mu A}%
%EndExpansion
$). The solid curve was taken without illumination whereas the dotted curve
was taken while the device was being illuminated by a monochromatic IR laser
light having a wavelength of $1550%
%TCIMACRO{\unit{nm}}%
%BeginExpansion
\operatorname{nm}%
%EndExpansion
$ and an effective power of $27%
%TCIMACRO{\unit{nW}}%
%BeginExpansion
\operatorname{nW}%
%EndExpansion
$. One notes that under illumination the resonance frequency substantially red
shifts while the damping rate surprisingly decreases. This is caused by the
fact that as the impedance of the HED is modified the RF current is repelled
out of the HED and redistributes in the ring \cite{Segev06a}, and consequently
the damping rate decreases. This measurement demonstrates frequency tuning by
monochromatic IR illumination, which is characterized by the parameter $\xi
Q\cong4.1$.

\begin{figure}
[ptb]
\begin{center}
\includegraphics[
height=3.4487in,
width=3.276in
]%
{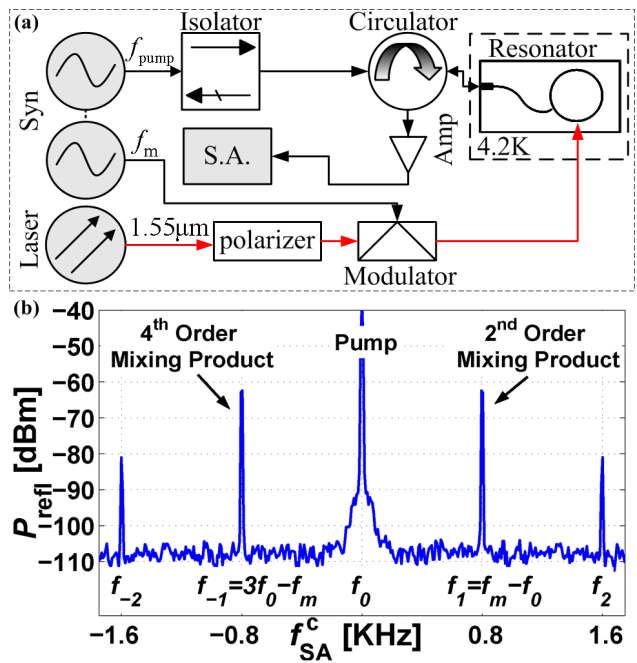}%
\caption{(Color online) $(\mathrm{a})$ Setup used for parametric excitation
measurement. $(\mathrm{b})$ The reflected power versus the measured frequency
$f_{\mathrm{SA}}$, centralized on $f_{0}$ ($f_{SA}^{c}=f_{SA}-f_{0}$).}%
\label{modulation setup}%
\end{center}
\end{figure}

Fast modulation of the resonance frequency is performed using the experimental
setup depicted in Fig. \ref{modulation setup}$(\mathrm{a})$. The resonator is
excited by a monochromatic pump tone, having a power of $-50.8$dBm, at
$f_{\mathrm{pump}}=f_{0}$. The optical power impinging on the HED, which has
an average value of $220$fW, is modulated at frequency $f_{\mathrm{m}%
}=2f_{\mathrm{pump}}+\Delta f\cong7.74%
%TCIMACRO{\unit{GHz}}%
%BeginExpansion
\operatorname{GHz}%
%EndExpansion
$, using a Mach-Zender modulator, driven by a second monochromatic signal,
phase locked with the first one. The frequency offset $\Delta f=800%
%TCIMACRO{\unit{Hz}}%
%BeginExpansion
\operatorname{Hz}%
%EndExpansion
$ is chosen to be much smaller than the resonance width $f_{0}/Q$. Note that
the laser power is approximately eight orders of magnitude lower than the
power used in the experimental approach proposed by Braggio \textit{et al}.
\cite{DCE_Braggio05}. Fig. \ref{modulation setup}$(\mathrm{b})$ shows the
reflected power off the resonator as a function of the measured spectrum
analyzer (SA) frequency $f_{\mathrm{SA}}$, centralized on $f_{0}$
$(f_{\mathrm{SA}}^{\mathrm{c}}=f_{\mathrm{SA}}-f_{0})$. It shows five
distinguished tones. The strongest one, labeled as $f_{0}$, is the reflected
spectral component at the frequency of the stimulating pump tone
$f_{\mathrm{pump}}$. The other four tones are found at frequencies
$f_{n}=f_{\mathrm{pump}}+n\Delta f$. For example, $f_{1}$ and $f_{-1}$ tones
results from second $(f_{1}=f_{\mathrm{m}}-f_{\mathrm{pump}})$ and fourth
$\ (f_{-1}=3f_{\mathrm{pump}}-f_{\mathrm{m}})$ order mixing between the
power-modulated optical signal and the driving pump tone respectively. No dc
bias current is employed in this measurement; however, the pump power is tuned
such that the HED is driven into a sub-critical region, close to a threshold
of a nonlinear instability
\cite{Baleegh06a,Baleegh06b,segev06b,segev06c,segev06d}.%
%TCIMACRO{\FRAME{ftbpFU}{3.276in}{3.4487in}{0pt}{\Qcb{(Color online)
%$(\QTR{rm}{a})$ Setup used for parametric excitation measurement.
%$(\QTR{rm}{b})$ The reflected power versus the measured frequency
%$f_{\QTR{rm}{SA}}$, centralized on $f_{0}$ ($f_{SA}^{c}=f_{SA}-f_{0}$%
%).}}{\Qlb{modulation setup}}{fig2.eps}{\special{ language "Scientific Word";
%type "GRAPHIC";  maintain-aspect-ratio TRUE;  display "ICON";
%valid_file "F";  width 3.276in;  height 3.4487in;  depth 0pt;
%original-width 13.8032in;  original-height 7.4105in;  cropleft "0";
%filename 'fig2.eps';file-properties "XNPEU";}}}%
%BeginExpansion
%EndExpansion

The results presented above demonstrate modulation of the resonance frequency
at the frequency of the primary parametric resonance. Furthermore, the
parametric gain threshold condition $\xi Q>1$ is achieved using optical
illumination having a fixed power. The main problem, which currently prevents
the achievement of parametric gain, is the relatively low photon flux that
impinges the HED. Due to losses along the optical path, especially the
expansion of the Gaussian beam from the tip of the fiber to the HED, the
largest photon flux, we currently manage to apply, is approximately $13$
photons per modulation cycle, at twice the resonance frequency. When taking
into account the quantum efficiency of the HED, which is probably lower than
$1\%$ \cite{HED_Korneev04}, and its effective area, which is probably smaller
than its printed area, we estimate that the optical power flux is about two
orders of magnitude lower than the threshold power. Focusing the laser beam in
future devices will allow overcoming this limitation.

Note however that, as was discussed above, the optical illumination employed
for parametric excitation results in some undesirable heating and consequently
an elevated noise. To study the noise properties of the system theoretically,
we employ a model in which the contribution of damping is taken into account
by adding a fictitious port coupled linearly to the resonator. The modes, in
both the feedline and in the damping port, are assumed to be in thermal
equilibrium at temperatures $T_{f}$ and $T_{d}$ respectively. Assume the case
where no RF signals are injected from noisy instruments at room temperature,
and the feedline is employed only for delivering the outgoing signal from the
resonator. In this case, employing an ultra-low noise cryogenic amplifier,
directly coupled to the feedline, may allow $T_{f}$ to be very close to the
base temperature of the refrigerator. On the other hand, $T_{d}$ might be much
higher due to the optical illumination. Assuming that damping in the resonator
occurs mainly in the illuminated section, one may assume that $T_{d}$ is close
to the temperature of that section. In general, however, modulating the
optical power drives that section out of thermal equilibrium, thus $T_{d}$
should be considered as an effective temperature characterizing the
non-equilibrium distribution. For the conditions appropriate for achieving
parametric gain one may assume that $T_{d}$ is close to the critical
temperature of the superconductor $T_{c}$.

When the resonance frequency varies in time according to Eq. (\ref{omega r})
the system acts as a phase sensitive amplifier \cite{Sqz_Movshovich90}. The
phase dependence can be studied by employing homodyne detection, namely, by
mixing the output signal, reflected off the resonator, with a local oscillator
at angular frequency $\omega_{0}$ (the parametric excitation is at frequency
$2\omega_{0}$) and with an adjustable phase $\phi$. We calculate the power
spectrum $S\left(  \omega,\phi\right)  $ of the homodyne detector output, in
the sub-threshold region, at angular frequency $\omega$. We find that
$S\left(  \omega,\phi\right)  $ is periodic in $\phi$ with period $\pi$. We
denote the minimum value as $S_{-}(\omega)$ (squeezed quadrature) and the
maximum one as $S_{+}(\omega)$ (amplified quadrature). The derivation is
similar to the one presented earlier in Ref. \cite{Squeezing_Yurke05}, thus we
only state here the final results%
\begin{align}
S_{\pm}(\omega)  &  =A\left\{  1-\frac{4\left(  \frac{1}{Q_{u}}\mp\xi\right)
}{Q_{f}\left[  \left(  \frac{2\omega}{\omega_{0}}\right)  ^{2}+\left(
\frac{1}{Q}\mp\xi\right)  ^{2}\right]  }\right\} \nonumber\\
&  +A\frac{4}{Q_{f}Q_{u}\left[  \left(  \frac{2\omega}{\omega_{0}}\right)
^{2}+\left(  \frac{1}{Q}\mp\xi\right)  ^{2}\right]  }\ , \label{S pm}%
\end{align}
where $A=\left(  1/2\right)  \coth\left(  \hbar\omega_{0}/2k_{B}T_{d}\right)
$, $Q_{u}$ is the unloaded quality factor of the resonator, which is related
to the loaded quality factor $Q$ by $1/Q=1/Q_{u}+1/Q_{f}$, where $Q_{f}$
characterizes the coupling between the resonator and the feedline.

Vacuum noise squeezing occurs when $S_{-}<0.5$. Consider as an example the
case where $\omega_{0}=2\pi\times5$ $%
%TCIMACRO{\unit{GHz}}%
%BeginExpansion
\operatorname{GHz}%
%EndExpansion
$ , $T_{f}=0.01$ $%
%TCIMACRO{\unit{K}}%
%BeginExpansion
\operatorname{K}%
%EndExpansion
$ , $T_{d}=10$ $%
%TCIMACRO{\unit{K}}%
%BeginExpansion
\operatorname{K}%
%EndExpansion
$, $Q_{f}=100$, $Q_{u}=2\times10^{4}$, $\omega=0$, and $\xi=0.01$. Using Eq.
(\ref{S pm}) one finds $S_{-}=0.2$. This example demonstrates that vacuum
noise squeezing can be achieved even when $\hbar\omega_{0}\ll k_{B}T_{d}$,
provided that the coupling to the feedline is made sufficiently strong.

In summary, we present preliminary experimental results which suggest that NbN
superconducting stripline resonators may serve as an ideal tool for studying
the DCE experimentally. Moreover we study theoretically the noise properties
of the system and find that vacuum noise squeezing may be achieved even when
the optical illumination employed for parametric excitation causes a
significant local heating.

This work was supported by the Israeli defense ministry, Israel Science
Foundation under grant 1380021, the Deborah Foundation and Poznanski
Foundation. One of the authors E.B. would especially like to thank Michael L.
Roukes for supporting the early stage of this research and for many helpful
conversations and invaluable suggestions.

%Just because of unusual number of tables stacked at end
\bibliographystyle{apsrev}
\bibliography{Bibilography}
%Produces the bibliography via BibTeX.

\end{document}